\documentclass [11pt,letterpaper]{JHEP3}
\pdfoutput=1
\usepackage{amsmath, bm, graphicx}
\usepackage{epstopdf}
\usepackage{amscd}
\usepackage{slashed}

\newcommand{\beq}{\begin{equation}}
\newcommand{\eeq}{\end{equation}}
\newcommand{\bea}{\begin{eqnarray}}
\newcommand{\eea}{\end{eqnarray}}
\newcommand{\nn}{\nonumber}

\def\Dslash{{\rlap{\raise 1pt \hbox{$\>/$}}D}}
\def \( {\left(}
\def \) {\right)}

\def\slashchar#1{\ensuremath{                               %
   \setbox0=\hbox{${}#1{}$}       
   \dimen0=\wd0                                 
   \setbox1=\hbox{/} \dimen1=\wd1               
   \ifdim\dimen0>\dimen1                        
      \rlap{\hbox to \dimen0{\hfil/\hfil}}      
      {}#1{}                                    
   \else                                        
      \rlap{\hbox to \dimen1{\hfil${}#1{}$\hfil}}   
      /                                         
   \fi}}       

\newcommand{\tr}{{\rm tr}}

\title{Can fermions save large N dimensional reduction?}
\author{
	{
	\def\href#1#2{#2}	
Paulo F. Bedaque\footnote{\email{bedaque@umd.edu}}, Michael I. Buchoff\footnote{\email{mbuchoff@umd.edu}}\, and Aleksey Cherman\footnote{\email{alekseyc@umd.edu}}\\
Maryland Center for Fundamental Physics\\
	Department of Physics, University of Maryland,
	College Park, MD 20742-4111
	}
	}
\author{
{
\def\href#1#2{#2}	
Roxanne P. Springer\footnote{\email{rps@phy.duke.edu}}\\
 Department of Physics, Duke University, Durham NC 27708
}
}
\abstract{This paper explores whether Eguchi-Kawai reduction for gauge theories with adjoint fermions is valid.  The Eguchi-Kawai reduction relates gauge theories in different numbers of dimensions in the large N limit provided that certain conditions are met.  In principle, this relation opens up the possibility of learning about the dynamics of 4D gauge theories through techniques only available in lower dimensions.  Dimensional reduction can be understood as a special case of large N equivalence between theories related by an orbifold projection. In this work, we focus on the simplest case of dimensional reduction, relating a 4D gauge theory to a 3D gauge theory via an orbifold projection.  A necessary condition for the large N equivalence between the 4D and 3D theories to hold is that certain discrete symmetries in the two theories must not be broken spontaneously.  In pure 4D Yang-Mills theory, these symmetries break spontaneously as the size of one of the  spacetime dimensions shrinks.   An analysis of the effect of adjoint fermions on the relevant symmetries  of the 4D theory shows that the fermions help stabilize the symmetries.    We consider the same problem from the point of view of the lower dimensional 3D theory and find that, surprisingly, adjoint fermions are not generally enough to stabilize the necessary symmetries of the 3D theory. In fact, a rich phase diagram arises, with a complicated pattern of symmetry breaking.  We discuss the possible causes and consequences of this finding.}
\begin{document}

\section{Introduction}
In the large N limit, gauge theories can have the remarkable property of volume independence. Under some circumstances, one or more of the dimensions of the spacetime in which the large N gauge theory lives can be shrunk, while a large set of observables remain unchanged.  In theories where this volume independence works all the way to zero size, large N volume independence opens up the prospect of relating four dimensional gauge theories to lower dimensional counterparts that can be studied by techniques only available in $D < 4$, for instance light-cone quantization or quantum mechanical variational methods.  Also, it may be possible to numerically simulate the lower-dimensional theories at a lower computational cost. 

The first version of the observation of volume independence in gauge theories is due to Eguchi and Kawai\cite{EK}, who argued that four dimensional lattice Yang-Mills theory in the large N limit is equivalent to a matrix model with no spacetime dimensions (a one plaquete model). However, there are conditions that need to be satisfied for the volume independence to hold. One of them is that no phase transitions occur as the size of the spacetime volume changes.  Unfortunately, in the case of pure Yang-Mills, a phase transition does occur as one of the dimensions is shrunk below a certain critical size\cite{quenched}. This is the finite temperature phase transition leading to deconfinement, where the center symmetry is broken. Thus, the reduction of the spacetime to one plaquete fails in this case. Some attempts were made to engineer constructions that would avoid this problem: the quenching procedure\cite{quenched}, the twisted model\cite{twisted} and, more recently, the double-trace deformation\cite{doubletrace,doubletraceShifmanUnsal}.  There are indications that the quenching\cite{quenched_fails} and the twisting\cite{twisted_fails} procedures fail to protect center symmetry near the continuum limit.

Parallel to these developments, a seemingly different kind of equivalence between large N gauge theories was discovered.  It was found that one can define `orbifold projections' that relate a `parent'  gauge theory to a `daughter' gauge theory, which is identical to the parent except that all degrees of freedom not invariant under some discrete symmetry of the parent theory are left out of the daughter theory.  In the large N limit, provided that certain conditions are met, there is then a class of `neutral' observables in the parent and daughter theories which are the same in both theories.   The parent and daughter theories are then termed to be `orbifold equivalent'.

Such orbifold equivalences were first discovered in string theory\cite{orbifold_string}, but were quickly realized to be purely field-theoretical effects\cite{orbifold_qft}.  Proofs of orbifold equivalence at the level of perturbation theory can be constructed by analyzing Feynman diagrams\cite{orbifold_qft,schmaltz}.  Orbifold equivalence and related techniques such as orientifold equivalence have mostly been used in the literature so far to relate supersymmetric and non-supersymmetric theories\cite{orbifold_review}.  

It was pointed out in ref.~\cite{seattle_vol_ind} that large N volume independence can be understood as a special case of orbifold equivalence. Volume independence can be seen as a volume reducing or a volume expanding transformation that leaves certain observables unaffected, and these transformations can be viewed as orbifold projections.  In viewing volume independence as a special case of orbifold equivalence, one is forced to work with the large N gauge theory defined on a lattice.  

For dimensional reduction, one starts with a theory which has one spacetime direction discretized on a lattice with $\Gamma$ points, and then projects by the discrete translation symmetry $\mathbb{Z}_\Gamma$.  Only the translation invariant fields survive the projection, so that the daughter theory is effectively defined in one dimension less than the parent.  In the volume expansion case\cite{MotherMoose,seattle_vol_ind}, when one performs `dimensional reconstruction', one starts with an $\mathop{\rm SU}(N\Gamma)$ gauge theory containing some unitary, adjoint scalars, and projects it by a discrete  $\mathbb{Z}_\Gamma$  symmetry
 acting on the color indices.  If the action of  $\mathbb{Z}_\Gamma$ on the different fields is chosen properly, the resulting daughter theory is a ``moose" or ``quiver"  $\mathop{\rm SU}(N)^\Gamma$ gauge theory with bi-fundamental scalar fields.  As is well known \cite{deconstruction}, such theories can be interpreted as theories with one extra dimension discretized on a lattice with $\Gamma$ sites. The role of the gauge links in the extra dimensions is played by the unitary scalars. The large N equivalence between neutral quantities in the parent and daughter theories then amounts to an equivalence between theories ``living" in different dimensions.
 
A non-perturbative proof of large N orbifold equivalence determines the necessary and sufficient conditions for its validity\cite{nonpert_proof}. These conditions must be satisfied in both the parent and daughter theories, and are i) unbroken center symmetry (confinement) and ii) unbroken $\mathbb{Z}_\Gamma$ symmetry, the symmetry used in the orbifold projection.  The violation of the first condition is what invalidates the original Eguchi-Kawai construction. Eguchi-Kawai reduction is valid as long as center symmetry would be unbroken, which is the case as long as the large N theory lives in a volume larger than a certain critical size of order $\Lambda_{QCD}^{-1}$. Numerical results support this statement \cite{finite_L}.  The proposed modifications of the Eguchi-Kawai construction such as the quenched and twisted models, as well as the double-trace deformed models, are attempts at preserving the symmetries necessary for large N volume independence (or orbifold equivalence) for volumes smaller than this critical size.

In this paper, we will focus on the simplest case of dimensional reduction, and relate a 4D gauge theory to a 3D gauge theory by an orbifold projection.   We will consider the effect of adjoint fermion matter fields on the realization of the discrete symmetries necessary for orbifold equivalence to hold so that large N dimensional reduction can work.  It has been suggested that adjoint fermions can prevent center symmetry breaking\cite{seattle_vol_ind} when a compactified spatial direction gets small in a 4D theory\footnote{Adjoint fermions only help protect spatial volume independence, since they must have periodic boundary conditions to protect center symmetry.  When fermions have antiperiodic boundary conditions, they do not help to protect center symmetry\cite{seattle_vol_ind}. }.  This suggestion was motivated by examining the behavior of the effective potential of a traced Wilson loop  wrapping a compact spatial direction in the $4D$ theory, which is a gauge-invariant order parameter for center symmetry breaking.  In the limit where the circumference of the compact spatial direction is small compared to $\Lambda_{QCD}^{-1}$,  the effective potential can be calculated perturbatively.  It was found that at one-loop order, the gauge boson contribution to the effective potential favors center symmetry breaking, while the contribution of each flavor of adjoint Majorana fermions is equal and opposite to that of the gluons, and favors the preservation of center symmetry.  The observation of this behavior in a 4D theory motivated the suggestion that 4D YM theories with adjoint fermions in the large $N$ limit have volume independence to zero size, so that one can construct dimensionally reduced theories, for instance 3D ones,  that are equivalent in the large N limit to 4D  theories due to orbifold equivalence\cite{seattle_vol_ind}. 

To be sure that large $N$ volume independence really holds to zero size in YM theories with adjoint fermions, however, it is necessary to check whether the necessary discrete symmetries are realized appropriately in the dimensionally reduced theory as well as in the 4D theory.  First, both the 3D and 4D theories should be in a confining phase, which is expected to be the case provided that both theories live in large enough spacetime volumes.  (Of course, as ref.~\cite{seattle_vol_ind} showed, the 4D theory stays in a confining phase even in small spatial volumes if adjoint fermions are present.) Next, the discrete $\mathbb{Z}_{\Gamma}$ translation symmetry along the compact direction of the 4D theory must not be broken spontaneously.  Fortunately, translation symmetry is not expected to break spontaneously in gauge theories like the ones that we are considering.\footnote{At non-zero chemical potential for baryon number in a large N theory, translation symmetry might break spontaneously due to the formation of  some kind of nuclear matter (for instance something like a Skyrme crystal of baryons) at a certain critical value of the chemical potential.}  Finally, it is crucial that the $\mathbb{Z}_{\Gamma}$ symmetry of the 3D theory must also not be broken spontaneously for large N dimensional reduction through orbifold equivalence to work.  It is this last question that we will focus on in this paper.

In section 2 we engineer a three dimensional theory that, upon a volume-expanding orbifold projection, leads to a latticized version of a four dimensional gauge theory with adjoint fermions. In section 3  we compute the relevant effective potential and discuss the spontaneous breaking of the $\mathbb{Z}_\Gamma$ symmetry used in the orbifold procedure. The minimization of this potential is discussed in section 4, and we find that, for most of the parameter space,  $\mathbb{Z}_\Gamma$ is broken and the large N equivalence fails. We then discuss the origin and consequences of this very surprising result.
 
\section{Dimensional reconstruction and fermions}
In this section, we construct a $3D$ theory which will be related to a $4D$ YM theory with adjoint fermions, and discuss its relevant discrete symmetries.  We then apply a volume-expanding orbifold projection to the $3D$ theory, and show that the projection produces a $4D$ theory with adjoint fermions.

\subsection{$D=3$ theory}
We will start with our $3D$ theory, which will be the parent from the point of view of the orbifold projection.  This theory is engineered in such a way as to generate, after a volume-expanding orbifold projection, a  daughter theory which is $4D$, $\mathop{\rm SU}(N)$ gauge theory coupled to one flavor of adjoint Dirac fermions.  The $4D$ theory will live on $\mathbb{R}^3 \times S^1$, and the circle will be discretized on a lattice with $\Gamma$ sites.  The fermions will have periodic boundary conditions on the circle, so that we are considering a spatial compactification in the 4D theory.

Since the $4D$ theory will have fermions and will be defined on a lattice, a naive discretization of the fermions will lead to fermion doubling in the continuum limit.  To prevent doubling, we will engineer the $3D$ theory to give rise to a $4D$ theory with Wilson fermions, as it turns out to be easiest to work out the orbifold prescription in this case.  As is well-known, the Wilson term breaks chiral symmetry in the $4D$ theory, and induces an additive renormalization of the fermion mass in the 4D theory.   This means that we also have to introduce a bare quark mass term, so that the $4D$ theory can have light fermions in the continuum limit if the bare quark mass is tuned appropriately against the Wilson term coefficient.

The $3D$ theory is a $\mathop{\rm \mathop{\rm SU}}(N\Gamma)$ gauge theory coupled to one adjoint unitary scalar and a pair of adjoint fermions.  We work in Minkowski space with the mostly minus metric.  The action of the $3D$ parent theory is
\bea\label{eq:3d}
 S_P &=&  \Gamma a\int d^3x\ {\rm tr} \left[ -\frac{1}{2 g^2}\mathbb{F}^2 +\frac{1}{2g^2a^2}|D_\alpha\phi|^2 
 + \bar\chi (i\rho^\alpha D_\alpha+m\tau^1)\chi \right.\nn\\
&&\left.\ \ \ \ \ \ \ \ \ \ \ \ \ \ \ \ \ \ \ \ \ \ \ \ \ \ \ \ \ \ \ \  -\frac{i}{2a} [\bar\chi , \phi^\dagger] \tau^3 \{\chi , \phi\}
 +\frac{r}{2a} [\bar\chi , \phi^\dagger] \tau^1 [\chi , \phi] \right],
 \eea 
 where $\alpha = 0,1,2$, $\rho^{\alpha}$ are the 3D Dirac matrices, the $D_{\alpha}$ are covariant derivatives in the adjoint representation,  $\chi$ is a doublet of two-component $3D$ adjoint Dirac fermions, $\tau^1$ and $\tau^3$ are the Pauli matrices $\sigma_1, \sigma_3$ acting on the $\chi$ flavor space, and $\phi \in SU(N\Gamma)$ is a unitary scalar field. The constants $g$, $a$, $\Gamma$ and $r$ will turn out to be, respectively,  the $4D$ dimensionless coupling constant, lattice spacing, the number of lattice points in the fourth direction, and the coefficient of the Wilson term.   To make the behavior of these fields once we move to the 4D theory more transparent, we have written the 3D action above in terms of fields with mass dimensions normalized according to the usual 4D conventions.
 
The kinetic term of the scalar $\phi$ will generate the gauge-kinetic terms in the $x_3$ direction after the orbifold projection.   Note that somewhat surprisingly, the would-be Wilson term (the term proportional to $r/a$ in the action) has the same mass dimension as the would-be $x_3$ kinetic term for the fermions (the term proportional to $i/a$) in the $3D$ action Eq.~(\ref{eq:3d}).  It will turn out that upon orbifold projection, these two terms will behave differently, and the Wilson term will become a  dimension 5 operator in the $4D$ theory, as it must.  
 
 We use the $D=2+1$ Dirac matrices defined as
 \bea
 \rho^0&=& \sigma^3,\nn\\
 \rho^1&=& i\sigma^2,\nn\\
 \rho^2&=& -i\sigma^1,
 \eea 
 where the $\sigma^{i}$ are the Pauli matrices.
 
 The action in Eq.~\eqref{eq:3d} has a discrete $\mathbb{Z}_\Gamma$ symmetry which acts on the fields as 
 \bea\label{eq:ZGamma_fields}
 \mathbb{A} &\rightarrow& \gamma \mathbb{A} \gamma^\dagger,\nn\\
 \chi &\rightarrow& \gamma \chi \gamma^\dagger,\nn\\
 \phi &\rightarrow& \omega \gamma \phi \gamma^\dagger,
 \eea 
 where $\omega=e^{\frac{2\pi i}{\Gamma}}$ and
 \beq
 \label{eq:ZGamma}
 \gamma = \left(\begin{array}{ccccc}
  \mathbf{1}_N &       &     &  &\\
   &\omega \mathbf{1}_N &     &  &\\
   &       &\omega^2 \mathbf{1}_N &      &\\ 
   &       &         &\ddots &\\
   &       &         &       &\omega^{\Gamma-1} \mathbf{1}_N 
 \end{array} \right),
  \eeq 
where $\mathbf{1}_N$ is the $N\times N$ identity matrix.  This $\mathbb{Z}_{\Gamma}$ symmetry will be the symmetry used to define the orbifold projection.

 \subsection{Orbifold projection to $D=4$}
The volume-expanding orbifold projection amounts to dropping all degrees of freedom in Eq.~\eqref{eq:3d} not invariant under Eq.~\eqref{eq:ZGamma_fields}. The only surviving components of the gauge and fermion fields are in the $N\times N$ diagonal blocks, while the surviving degrees of freedom of the scalar field are in the $N\times N$ one-off-diagonal blocks.
 
 \bea\label{eq:projected}
 \mathbb{A}\rightarrow
 \left(\begin{array}{ccccc}
 \mathbb{A}_1 &       &     &  &\\
              &\mathbb{A}_2 &     &  &\\
              &       &\mathbb{A}_3 &      &\\ 
              &       &         &\ddots &\\
              &       &         &       &\mathbb{A}_\Gamma
 \end{array} \right), \ \ \ 
 \mathbb{\chi}\rightarrow
 \left(\begin{array}{ccccc}
 \mathbb{\chi}_1 &       &     &  &\\
              &\mathbb{\chi}_2 &     &  &\\
              &       &\mathbb{\chi}_3 &      &\\ 
              &       &         &\ddots &\\
              &       &         &       &\mathbb{\chi}_\Gamma
 \end{array} \right),\ \ \
 \phi\rightarrow
 \left(\begin{array}{ccccc}
 &\phi_1 &       &     &  \\
 &             &\phi_2 &     &  \\
 &             &       &\phi_3 &      \\ 
 &             &       &         &\ddots \\
 \phi_\Gamma&             &       &         &       
 \end{array} \right).
 \eea 
 
  The action of the daughter theory is, up to a factor, the orbifold-projected action of the parent ($3D$) theory $S_3$:
 \beq
 S_3[\mathbb{A}, \phi, \chi] \ \ \ \Longrightarrow \ \ \ S_{D}[\mathbb{A}_1, \cdots \mathbb{A}_\Gamma, \phi_1\cdots \phi_\Gamma, \chi_1\cdots \chi_\Gamma]
  \equiv \frac{1}{\Gamma}\sum_{n=0}^{\Gamma-1} 
  S_P[ \gamma^n \mathbb{A}\gamma^{n\dagger},  \gamma^n \chi \gamma^{n\dagger}, \omega^n \gamma^n \phi\gamma^{n\dagger} ]
 \eeq 

When valid, the orbifold equivalence states that,  to leading order in $1/N$, the correlators of {\it neutral} operators in the parent (that is, operators invariant under the the $\mathbb{Z}_\Gamma$ transformation in Eq.~\eqref{eq:ZGamma_fields}) agree with the correlators in the daughter theory.  An example of a neutral operator in the $3D$ theory is $\tr \phi^{\Gamma}$, which is a Wilson loop wrapping the compact $S^1$ direction from the point of view of the $4D$ theory.

In the case of the action $S_P$ in Eq.~\eqref{eq:3d} we have the daughter theory action
  
 \bea\label{eq:4dlattice}
 S_D &=&
 \int d^3x \ a\sum_{i=1}^\Gamma{\rm Tr}\left[ -\frac{1}{2g^2}\mathbb{F}^i_{\alpha\beta}\mathbb{F}^{i\alpha\beta}
 + \frac{1}{a^2g^2}|D_\alpha\phi_{i}|^2 \right.
 + \bar\chi_i (i\rho^\alpha D_\alpha+m\tau^1)\chi_i \\
 & & \ \ \ \ \ \ \ \ \ \ \ \ \left.
   -\frac{i}{2a} (\bar\chi_i  \phi_i^\dagger - \phi_i^\dagger \bar\chi_{i+1}   ) \tau^3  (\chi_i  \phi_i + \phi_i \chi_{i+1}   )
 +\frac{r}{2a} (\bar\chi_i  \phi_i^\dagger - \phi_i^\dagger \bar\chi_{i+1}   ) \tau^1  (\chi_i  \phi_i - \phi_i \chi_{i+1}   ) \right],\nn
 \eea where 
 \bea
 D_\alpha\phi_{i} &=& \partial_\alpha\phi_{i} +i(\mathbb{A}_\alpha^{i} \phi_{i}-\phi_{i}\mathbb{A}_\alpha^{i+1} ),\nn\\
  D_\alpha\chi_{i} &=& \partial_\alpha\chi_{i} +i[\mathbb{A}_\alpha^{i} ,\chi_{i}] .
 \eea 
 The action in Eq.~\eqref{eq:4dlattice} is the action for a $4D$ gauge theory discretized in the compact $x^3$ direction, with the unitary scalar $\phi_i$ as the $i$-th link variable in the $x^3$ direction. In fact, writing 
 $\phi_i = e^{ia\mathbb{A}_3(i)}$  we see that the kinetic term for $\phi$  is a discretized version of $\mathbb{F}_{\alpha 3}$:
 \bea
 D_\alpha\phi_{i} &\simeq& \partial_\alpha\phi_{i} + i \mathbb{A}_\alpha^{i}\phi_{i}- i \phi_{i}\mathbb{A}_\alpha^{i+1} \nn \\
 &\simeq& ia\left( \partial_\alpha \mathbb{A}_3^{i}-\frac{\mathbb{A}_\alpha^{i+1}-\mathbb{A}_\alpha^{i}}{a}
 +i(\mathbb{A}_\alpha^{i}\mathbb{A}_3^{i}-\mathbb{A}_3^{i}\mathbb{A}_\alpha^{i+1})\right)
 \simeq ia\mathbb{F}_{\alpha 3}^{i},
 \eea  
 Similarly
 \beq
 \sum_{i=1}^\Gamma (\bar\chi_{i+1} \tau^3 \phi_i^\dagger \chi_i\phi_i -\bar\chi_i \tau^3 \phi_i \chi_{i+1}\phi_i^\dagger)
 \simeq 2ia \bar\chi i\tau^3 D_3\chi.
 \eeq and
 \beq
  \sum_{i=1}^\Gamma(\bar\chi_{i+1} \tau^3 \phi_i^\dagger \tau^1\chi_i\phi_i + \bar\chi_i \tau^3 \phi_i \tau^1\chi_{i+1}\phi_i^\dagger - 2 \bar\chi_i\tau^1\chi_i)
  \simeq -a^2 \bar\chi \tau^1(D_{3})^2\chi.
 \eeq We now define the four-component $4D$ Dirac spinor $\psi$ by
 \beq\label{eq:Dirac}
 \psi = \left(\begin{matrix}
         \chi^1\\
         \sigma^3 \chi^2
         \end{matrix}\right),
 \eeq  
 and the $4D$ Dirac matrices in the chiral basis in terms of the $3D$ Dirac matrices as
 \beq
 \gamma^\mu = \begin{pmatrix}
         0 & \bar\sigma^\mu\\
         \sigma^\mu & 0 
         \end{pmatrix},
 \eeq 
 with $\sigma^\mu=(1,\vec{\sigma})$ and $\bar\sigma^\mu=(1,-\vec{\sigma})$.  Essentially, the $3D$ Dirac spinors become $4D$ Weyl spinors in the orbifold construction.

 With the identifications above, we see that in the small $a$ limit the action in Eq.~\eqref{eq:4dlattice} becomes
 \beq\label{eq:4dcontinuum}
  S_D = \int d^4x\ {\rm tr} \left[ -\frac{1}{2g^2}\mathbb{F}^2 
 + \bar\psi (i\gamma^\mu D_\mu+m)\psi 
 -\frac{ra}{2} D_3\bar\psi D_3\psi
 +\mathcal{O}(a^2)
 \right].
 \eeq 

As promised, the orbifold projection takes a theory defined in $3D$ and generates a $4D$ theory, with the Wilson term becoming an irrelevant dimension 5 operator, as it must.  The $4D$ theory has gauge group $\mathop{\rm \mathop{\rm SU}}(N)$, in contrast to the $\mathop{\rm SU}(N\Gamma)$ gauge group of the $3D$ parent theory. The spatial extent of the reconstructed $x^3$ direction equals $L=\Gamma a$ and thus can be adjusted by tuning the parameters of the $3D$ theory. 
 
It is possible to adapt the construction above to accommodate $4D$ theories with  Majorana fermions, such as $\mathcal{N}=1$ Super Yang-Mills theory, which has one flavor of massless Majorana adjoint fermions\footnote{Supersymmetry would be broken by the lattice discretization.}.  To do this we can impose the condition $\sigma^2 (\chi^1)^* =\sigma^3\chi^2$ in Eq.~\eqref{eq:Dirac}, so that the 4D Dirac spinor written in terms of 3D spinors has the form 
 \beq
 \psi=\left(\begin{matrix}
         \chi^1\\
         \sigma^2 \chi^{1*}
         \end{matrix}\right),
 \eeq
 and becomes a $4D$ Majorana spinor. Setting $\sigma^2 (\chi^1)^* =\sigma^3\chi^2$, the action $S_3$ becomes 
 
 \bea
 S^{\mathrm{Majorana}}_{3} &=&  \Gamma a\int d^3x\ {\rm tr} \left[ -\frac{1}{2 g^2}\mathbb{F}^2 +\frac{1}{2g^2a^2}|D_\alpha\phi|^2 
 + 2\bar\chi i\rho^\alpha D_\alpha\chi +m(\chi^\dagger \sigma^2\chi^*+\chi^T \sigma^2\chi)\right.\\
 &&\ \ \ \ \ \ \left.
 -\frac{i}{a} ([\chi^* , \phi^\dagger]  \{\chi^* , \phi\}-[\chi^* , \phi^\dagger]  \{\chi^* , \phi\})
 -\frac{r}{2a} ([\chi^* , \phi^\dagger] [\chi^* , \phi] 
               +[\chi , \phi^\dagger]  [\chi , \phi]) \right],\nn
 \eea 
 where now $\chi$ does not carry any flavor indices.


\section{Large N equivalence and the effective potential}
\label{sec:EffPot}
For the $3D$ theory to be orbifold equivalent to the $4D$ theory in the large N limit, two basic conditions must be met.  First, both theories must be in their confined phases, and second, the symmetry that defines the orbifold projection, in our case the  $\mathbb{Z}_\Gamma$ symmetry in Eq.~\eqref{eq:ZGamma_fields}, must remain unbroken in the parent theory.   We expect the $3D$ theory to be in a confining phase as long as it is defined in a large enough 3D volume.  The $4D$ theory is expected to be in a confined phase when $L = \Gamma a$ is large enough, and also when $L < \Lambda_{QCD}^{-1}$ because of the presence of the adjoint fermions\cite{seattle_vol_ind}.    The crucial question that will determine whether large N orbifold equivalence between the two theories is valid is whether or not the $\mathbb{Z}_{\Gamma}$ symmetry of the $3D$ is spontaneously broken.

A set of {\it gauge invariant} order parameters for the $\mathbb{Z}_{\Gamma}$ symmetry is comprised of the expectation values of the traces of powers of $\phi$,  $\langle \tr \phi^k \rangle$, for $k=1, \cdots, \Gamma$. The information encoded in the $\Gamma$ quantities $\langle \phi^k \rangle$ is also encoded in the eigenvalues of $\phi$ (but not in their ordering, which is a gauge-dependent quantity).  We will investigate whether $\mathbb{Z}_\Gamma$ is spontaneously broken by calculating the effective potential for the eigenvalues of $\phi$. Actually, we will do a  more general calculation. We can diagonalize $\phi$ through a gauge transformation $\Omega$ as $\phi=\Omega\phi_0\Omega^\dagger$, where $\phi_0$ is diagonal:

\beq\label{eq:phi0}
 \phi_0=
 \left(\begin{array}{ccccc}
 e^{i\varphi_1} &                    &         &                                \\
                     & e^{i\varphi_2}&         &                                \\ 
                     &                    &          &\ddots &                 \\
                     &                    &          &          &e^{i\varphi_{N\Gamma}} 
 \end{array} \right).
 \eeq 
 In order to test for $\mathbb{Z}_\Gamma$ breaking, it would be enough to consider  a $\phi_0$ with only $\Gamma$ distinct eigenvalues, so that $\phi_0$ has a `block diagonal' form if they are properly ordered. We will, however, consider the more general case where all eigenvalues can be different, and test for the breaking of the full $\mathbb{Z}_{N\Gamma}$ symmetry defined by Eq.~\eqref{eq:ZGamma_fields} with 
 $\omega=e^{2\pi i/N\Gamma}$ and
 
\beq
 \gamma = \left(\begin{array}{ccccc}
  1 &       &     & \\
   &\omega  &     &  &\\
         &         &\ddots &\\
  &         &       & \omega^{N\Gamma-1}  
 \end{array} \right).
 \eeq 
 
The effective potential for the eigenvalues of $\phi$ will then determine how the $\mathbb{Z}_{N\Gamma}$ symmetry and its $\mathbb{Z}_{\Gamma}$ subgroup are realized.  Of course, the computation of the effective potential for $\langle tr \phi^k\rangle$, or equivalently for the distribution of the eigenvalues of $\phi$, is in general a difficult non-perturbative problem. However,  we expect the relevant scale for the renormalization of $g$ to be the size $L$ of the ``extra" dimension\footnote{The only other scale available in the 3D theory is $a\ll L$.}, as this is the case for gauge theories in $\mathbb{R}^3\times S^1$. For $L=a\Gamma\ll \Lambda_{QCD}^{-1}$, where $\Lambda_{QCD}$ is the strong scale of the $4D$ theory, we expect a semiclassical approximation to be valid, and we will compute the one-loop effective potential for the $\mathbb{Z}_{N\Gamma}$ symmetry.

The one-loop effective potential will be given by the small (quadratic) fluctuations around $\phi=\phi_0$. A perturbative calculation like this requires gauge fixing, and we will use $R_\xi$ gauge. We parametrize the $\phi$ field by $\phi=e^{i\mathbb{G}}\phi_0e^{-i\mathbb{G}}$, where the $\mathbb{G}$'s are the would-be Goldstone bosons.  
 
Gauge-fixing amounts to the addition to the lagrangian of
 \beq
 \mathcal{L} \rightarrow \mathcal{L}-\frac{1}{2}FF^\dagger+\bar c(\frac{\delta F_\alpha}{\delta\alpha})c, 
 \eeq where $\bar c,c$ are the Faddeev-Popov ghosts. We choose the gauge fixing function $F$ to be 
 \beq
 F=\frac{1}{g\sqrt{\xi}} [\partial.\mathbb{A} - \frac{\xi}{a^2}\mathbb{G},\phi_0].
 \eeq  
 
After gauge fixing, the part of the lagrangian quadratic in $\mathbb{A}, \partial\phi_0, \mathbb{G}$ and bilinears in $\bar c,c$, is
 \bea
 \frac{1}{\Gamma a}\mathcal{L}_{\rm fixed}^{quadratic} &=&
 -\frac{1}{2g^2}{\rm tr}\left(2\partial_\alpha \mathbb{A}_\beta \partial^\alpha \mathbb{A}^{\beta} - 2(\partial.\mathbb{A})^2 +\frac{1}{a^2}[\mathbb{A},\phi_0][\mathbb{A},\phi_0^\dagger]
 -\frac{1}{\xi}[\partial.\mathbb{A},\phi_0][\partial.\mathbb{A},\phi_0^\dagger]\right)\nn\\
 &&-\frac{1}{2g^2a^2}{\rm tr}\left([\partial \mathbb{G},\phi_0][\partial \mathbb{G},\phi_0^\dagger]
  -\frac{\xi}{a^2}[ \mathbb{G},\phi_0][ \mathbb{G},\phi_0^\dagger]\right)\nn\\ 
  &&+{\rm tr}\bar c (-\partial^2 -\frac{\xi}{a^2}) [c,\phi_0]+\frac{1}{2g^2a^2}{\rm tr}|\partial\phi_0|^2\nn\\
 && + {\rm tr}\bar\chi (i\rho^\alpha D_\alpha+m\tau^1)\chi 
 -\frac{i}{2a} [\bar\chi , \phi_0^\dagger] \tau^3 \{\chi , \phi_0 \}
 -\frac{r}{2a} [\bar\chi , \phi_0^\dagger] \tau^1 [\chi , \phi_0]
 .
 \eea 
 The would-be Goldstone bosons in $\mathbb{G}$ pick up a gauge dependent mass $m_G^2\sim \xi/a^2$, the ghosts pick up a mass $m_c^2\sim \xi/a^2$ and the longitudinal components of the $A$ that gets higgsed also pick up a mass $m_A^2\sim \xi/a^2$. The transverse components of the higgsed $A$'s (we shall call them $W$'s) pick up a real mass $m_W\sim 1/a$ and the fermions $\chi$ also get a mass $m_{\chi} \sim m+1/a$.	 The precise values of these masses depend on the value of $\phi_0$. To find them, let us parametrize the adjoint Hermitian fields as $\mathbb{A} = A^a \lambda^a$, with ${\rm tr}\lambda^a\lambda^b = \frac{1}{2}\delta^{ab}$, and similarly for $\mathbb{G}, \chi, \bar c$ and$\; c$.   We need to evaluate terms such as $ {\rm tr} (\lambda^a \phi^\dagger_0 \lambda^b \phi_0)$, which is easiest to do by considering the decomposition of $SU(N \Gamma)$ into its $SU(2)$ subgroups. It is not hard to show that if $\lambda_a, \lambda_b$ belong to different $SU(2)$ subgroups, or to the Cartan subalgebra, they do not make a $\phi_0$ dependent contribution to the effective potential.  This means that we only need to consider a sum over $SU(2)$ subgroups with $\lambda_a, \lambda_b$ belonging to the same subgroup.
 
To show how this works, we will temporarily work with $N\Gamma=2$; the generalization to a general $SU(N \Gamma)$ gauge group is straightforward and will be done at the end of the calculation. 

We will need the following traces: 
  \bea\label{eq:matrix_elements}
 {\rm tr} (\lambda^a \phi^\dagger_0 \lambda^b \phi_0) &=&  
 \begin{pmatrix}
 \frac{1}{2}\cos(\phi_1-\phi_2)  &  \frac{1}{2}\sin(\phi_1-\phi_2)&0 \\
 -\frac{1}{2}\sin(\phi_1-\phi_2)& \frac{1}{2}\cos(\phi_1-\phi_2)   & 0   \\
 0&   0  &  \frac{1}{2}   \\
 \end{pmatrix},\nn\\
 {\rm tr} ([\lambda^a, \phi^\dagger_0]\{ \lambda^b, \phi_0\}) &=& 
 {\rm tr} (\lambda^a \phi^\dagger_0 \lambda^b \phi_0-\lambda^a \phi_0 \lambda^b) =
 \begin{pmatrix}
 0  & \sin(\phi_1-\phi_2)&0 \\
 \sin(\phi_1-\phi_2)& 0   & 0   \\
 0&   0  &  0   \\
 \end{pmatrix},\\
 {\rm tr} ([\lambda^a, \phi^\dagger_0][ \lambda^b, \phi_0]) &=& 
 {\rm tr} (\lambda^a \phi^\dagger_0 \lambda^b \phi_0+\lambda^a \phi_0 \lambda^b \phi^\dagger_0-2\lambda^a\lambda^b) =
 \begin{pmatrix}
 \cos(\phi_1-\phi_2)-1  &  0&0 \\
0& \cos(\phi_1-\phi_2)-1   & 0   \\
 0&   0  &  0   \\
 \end{pmatrix}.\nn
 \eea 
 
Using Eq.~(\ref{eq:matrix_elements}) we see that only the $1$ and $2$ components of the fields are affected by $\phi_0$ and pick up a $\phi_0$ dependent mass, so we will drop the $3^{rd}$ component from now on. We arrive at
 \bea
 \frac{1}{\Gamma a}\mathcal{L}_{\rm quadratic}^{\rm fixed} &=&
 -\frac{1}{2g^2}\sum_{i=1,2}{\rm tr}\left(\partial_\alpha A^i_\beta \partial^\alpha A^{i\beta} - (\partial.A^i)^2 +\frac{1}{a^2}(\cos(\varphi_1-\varphi_2)-1)A^iA^i  \right.\nn\\
 &&\left. \ \ \ \ \ \ \ \ \ \ \ \ \ \ \ \ \ \ \ \ \ \ -\frac{1}{\xi}(\cos(\varphi_1-\varphi_2)-1)\partial.A^i\partial.A^i\right)\nn\\
 &&-\frac{1}{2g^2a^2}(\cos(\varphi_1-\varphi_2)-1)\sum_{i=1,2}\left(\partial G^i\partial G^i 
  -\frac{\xi}{a^2}G^i G^i\right)\nn\\ 
  &&+(\cos(\varphi_1-\varphi_2)-1)\sum_{i=1,2}\bar c^i (-\partial^2 -\frac{\xi}{a^2}) c^i   \nn\\
 && + \sum_{i=1,2}\bar\chi^i (i\rho^\alpha D_\alpha+m\tau^1)\chi^i 
 -\frac{i}{2a} \sin(\varphi_1-\varphi_2)(\bar\chi^1\tau^3\chi^2-\bar\chi^2\tau^3\chi^1)   \nn\\
 && \  \ \ \ \ \ \ \ \ \ \ \ \ \ \ \ \ \ \ \ \  \ \ \ \ \  \ \ \ \ \ \ \ \ \ \ \ \ \ +\frac{r}{2a}(\cos(\varphi_1-\varphi_2)-1) \bar\chi^i \tau^1\chi^i. 
 \eea where the superscripts in $\chi^{1,2}$ denote adjoint color, not flavor.

 The propagator for the color components $1$ and $2$ of the gauge fields can now be shown to be
 \beq
 iD_{\mu\nu}(k) = \frac{1}{k^2-\frac{1-\cos(\varphi_1-\varphi_2)}{a^2}}\left( g_{\mu\nu}-\frac{k_\mu k_\nu}{k^2}\right)
 +\frac{k_\mu k_\nu}{k^2} \frac{\xi}{(1-\cos(\varphi_1-\varphi_2))(k^2-\frac{\xi}{a^2})}.
 \eeq 
 The two physical, transverse degrees of freedom have a $\xi$-independent mass. The unphysical, longitudinal component has a $\xi$-dependent mass and wave function normalization. This mass (and wave function normalization) is identical to that of the would-be Goldstones and of the ghosts. Counting the unphysical degrees of freedom (d.o.f.) we have $2$ gauge d.o.f.,  $2$ Goldstone d.o.f., and $2$ ghost d.o.f. As the ghosts contribute to the effective potential with an additional factor of $-2$, the contribution of the unphysical degrees of freedom to the effective potential cancels out. We are left with the physical degrees of freedom: 4 gauge field d.o.f.  (from two adjoint colors and two polarizations) and 4 fermion d.o.f.  (from two adjoint colors and two ``flavors"). The Higgs field contribution does not depend on $\phi_0$, so we can ignore it. The one-loop effective potential will depend on $\phi_0$ only through the value of these physical masses and will, consequently, be $\xi-$independent. 
 
 The bosonic contribution to the effective potential can now be shown to be
 \beq\label{eq:integral}
 4 \int \frac{d^3k}{(2\pi)^3} \log\left(k^2-\frac{2}{a^2}\sin^2\left(\frac{\varphi_1 - \varphi_2}{2}\right)\right)\nn\\
 =
 \frac{4}{6\pi a^3} \left[2\sin^2\left(\frac{\varphi_1 - \varphi_2}{2}\right)\right]^{3/2},
 \eeq 
 where we have used the dimensionally regulated and minimally subtracted integral

 \beq\label{eq:integral}
\pm \int\frac{d^3k}{(2\pi)^3} \log(k^2-M^2) = \mp\frac{i}{6\pi}(M^2)^{3/2}.
 \eeq
At this point, one might worry that the integral above might have an infrared divergence for some distributions of eigenvalues, since the $M^2$ of the bosons vanishes for $\varphi_1=\varphi_2$.  This is a serious issue in a full Eguchi-Kawai reduction to a $D=0$ one-plaquette model, where the contribution of the gauge bosons to the one-loop effective potential becomes singular, and one must examine higher order corrections to the effective potential to make reliable conclusions\cite{ZeroModes}\footnote{We thank Larry Yaffe for bringing this issue to our attention.}. However,  the integral above for $D=1,2,3$ is IR-finite and well-behaved as $M \rightarrow 0$, so that this issue does not arise for $D>0$.

 The dispersion relation for the fermions can be found in  a similar way, where we continue to work with the $N\Gamma=2$ example for clarity. We find
 \beq
 \frac{1}{\Gamma a}\mathcal{L}^\chi_{quadratic} =\int \frac{d^3k}{(2\pi)^3}\frac{1}{2}(\bar\chi^1 \ \bar\chi^2)
        \underbrace{
        \begin{pmatrix}
            \slashed{k} +(m+\frac{r}{a}(1-\cos))\tau^1  &   -\frac{i}{a}\sin \tau^3   \\
             \frac{i}{a}\sin \tau^3   & \slashed{k} +(m+\frac{r}{a}(1-\cos))\tau^1
        \end{pmatrix}
        }_{M}
        \begin{pmatrix}
        \chi^1\\
        \chi^2
        \end{pmatrix}       
 \eeq 
 where we define the shorthand notation $\cos \equiv \cos(\varphi_1-\varphi_2)$ and $\sin \equiv \sin(\varphi_1-\varphi_2)$ and the indices on $\chi^{1,2}$ refer to color space. 
 The fermion matrix can be written as
 \beq
 M =(\slashed{k} +(m+\frac{r}{a}(1-\cos)\tau^1) \mathbb{I} + \frac{1}{a}\sin \tau^3 \Sigma^2
 \eeq with $\mathbb{I}$ and $\Sigma^2=\sigma^2$ acting on color space.  The fermion contribution to the effective potential is given by the Tr Log($M$) and can be calculated as
 \bea
 {\rm Tr \; Log}M &=& \int \frac{d^3k}{(2\pi)^3} {\rm tr}_{color}{\rm tr}_{flavor}{\rm tr}_{spin} \log M\nn\\
 &=&
 4 \int \frac{d^3k}{(2\pi)^3} \log\left(k^2-((m+\frac{r}{a}(1-\cos))^2+ (\frac{1}{a}\sin)^2)\right)\nn\\
 &=&
 \frac{2}{3\pi a^3} \left[(ma+2r\sin^2(\frac{\varphi_1-\varphi_2}{2}))^2+ \sin^2(\varphi_1-\varphi_2)\right]^{3/2}.
 \eea 
 Above we used relations like
 \beq
 {\rm tr}_{spin} \log(\slashed{k}+A) =  \log(k^2-A^2).
 \eeq

Adding up the gauge boson and fermion contributions to the effective potential we find
\begin{equation*}
V^{{\textrm SU(2)} }_{\mathrm{eff}}(\phi_0) =  \frac{1}{6\pi a^3} \left[4 \sqrt{8}|\sin(\frac{\varphi_1-\varphi_2}{2})|^3  -8\left|(ma+2r\sin^2(\frac{\varphi_1-\varphi_2}{2}))^2+ \sin^2(\varphi_1-\varphi_2)\right|^{3/2}\right].
\end{equation*}

It is straightforward to generalize the calculation outlined above to the  $\mathop{\rm SU}(N\Gamma)$ case, with the result that
\beq
\label{eq:Veff}
V_{\mathrm{eff}}(\phi_0) =  \frac{4}{3\pi a^3} \sum_{i<j=1, ..., N\Gamma}\left[\sqrt{2}|\sin(\frac{\varphi_i-\varphi_j}{2})|^3 
-  \left|(ma+2r\sin^2(\frac{\varphi_i-\varphi_j}{2}))^2+ \sin^2(\varphi_i-\varphi_j)\right|^{3/2}\right].
\eeq 
This is the main result of this section.

\section{The phase diagram for $\mathbb{Z}_{N \Gamma}$ symmetry}
\label{sec:PhaseDiagram}
A few observations about the effective potential in Eq.~\eqref{eq:Veff} can help us gain insight into the result. First, let us consider  the purely bosonic case. In this case the effective potential reduces  to 
 \beq\label{eq:Veff_gaugeonly}
V_{\mathrm{eff}}(\phi_0) =  \frac{4\sqrt{2}}{3\pi a^3} \sum_{i<j=1, ..., N\Gamma} |\sin(\frac{\varphi_i-\varphi_j}{2})|^3 .
\eeq 
It is clear that the effective potential is minimized when all of the eigenvalues are degenerate and $\langle \tr \ \phi_0^k\rangle \neq 0$. This ground state breaks $\mathbb{Z}_{N\Gamma}$ maximally, down to nothing. In particular, the $\mathbb{Z}_\Gamma$ symmetry used to define the orbifold projection is spontaneously broken, which implies that the equivalence between the large N dynamics of the $3D$ and $4D$ theories is lost. This conclusion agrees with the standard belief that the Eguchi-Kawai reduction for pure YM breaks down if one (or more) dimensions are shrunk down to one plaquete.

Let us now look at the fermion contribution to the effective potential.  First, note the surprising fact that the Wilson term coefficient shows up in the effective potential, and does not disappear in the small $a$ limit.  This can be traced back to the fact that the 3D progenitor of the 4D Wilson term shows up at the same order as everything else in the 3D action, even though the Wilson term is an irrelevant operator in the $4D$ theory.  The effective potential of the 3D theory thus depends strongly on the value of the Wilson term coefficient.

Next, note that the fermion mass will be additively renormalized from its bare value by loop contributions due to the couplings of the fermions to the gauge field. Those effects are higher order in $g^2$ and are not reflected in Eq.~(\ref{eq:Veff}).  The renormalization of the mass would show up at two-loop order in the calculation of the effective potential.  Thus, at the order to which we are working, the parameter $m$ in Eq.~(\ref{eq:Veff}) can be viewed as the ``fermion mass" (with the obvious issues relating to the proper definition of fermion mass in a confining theory applying).

Now we must ask what region in the parameter space of the $3D$ theory can correspond to a continuum $4D$ theory, assuming that the orbifold equivalence is valid.  The continuum limit in the $4D$ theory corresponds to sending $a\rightarrow 0, \Gamma \rightarrow \infty$ in such a way that $\Gamma a = L$ is fixed. This means that the lattice spacing $a$ must be much smaller than any other length scale in the problem, and in particular:
\beq
m a \ll 1 \;.
\eeq
This means that we should only expect to have our $3D$ theory to be equivalent to a continuum $4D$ theory with fermions in a region in parameter space where $ m a \ll 1 $.

The  general effect of the fermion contribution to $V_{\mathrm{eff}}$ naively seems to counteract the symmetry breaking effect of the gauge fields, as the bosonic and fermionic contributions have opposite signs. For instance, for the special case $ma=0$ and $r=1$, where we might hope that the equivalence should work, the potential between any two eigenvalues is minimized when they are maximally separated ($\phi_i - \phi_j = \pm \pi$). That does not mean, however, that the $\mathbb{Z}_{N\Gamma}$ symmetric distribution of eigenvalues (where $e^{i\phi_n} =e^{i2\pi i n/N\Gamma} , n=1,\cdots,N\Gamma $) is favored, as the symmetric configuration does not minimize $V_{\mathrm{eff}}$ for {\it each pair} of eigenvalues. $V_{\mathrm{eff}}$ can, instead, be minimized by having half of the eigenvalues equal $1$ and the other half equal to $-1$. The energy lost by having pairs of identical eigenvalues is more than compensated for by having even more pairs with the maximal separation.  As a result, the $\mathbb{Z}_{\Gamma}$ symmetry used in the orbifold projection breaks spontaneously, and we see that surprisingly large N orbifold equivalence is \emph{not} valid for $m a = 0, r=1$, and the $3D$ theory is not equivalent to a continuum $4D$ theory.

\begin{figure}[t] 
\centering
\includegraphics[scale=0.45]{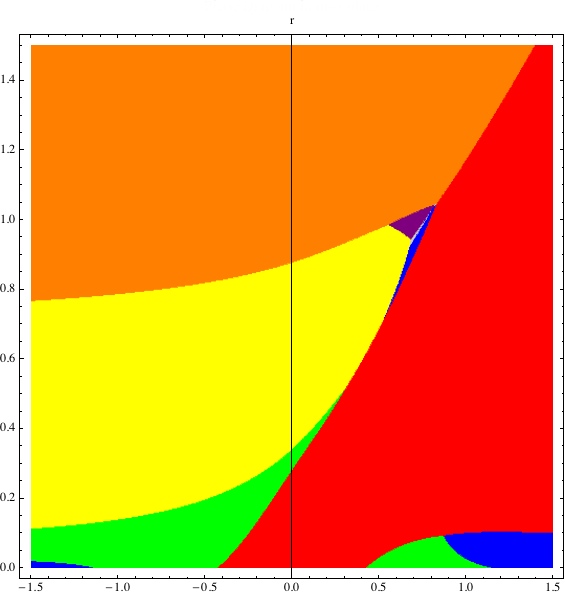}
\caption{Phase plot of one-loop effective potential with axes $r$ vs. $am$ in the large $N\Gamma$ limit.  Configurations of the $N\Gamma$ eigenvalues  distributed at $1, \ldots,i\ldots, N \Gamma$ points evenly spaced around the circle were tested against each other, and the configuration with the lowest energy was determined at each point in the plot above.  We refer to the configuration with the eigenvalues distributed on $i$ points on the circle as the $Z_{i}$ configuration in the plot legend above.  The purple region is the region where the full $Z_{N \Gamma}$ symmetry is unbroken.}
\label{fig:1}
\end{figure}

For generic values of $ma$ and $r$, the minimization of  $V_{\mathrm{eff}}$ is a complex problem and better dealt with numerically. The result of such a minimization is summarized in Fig.~\ref{fig:1}. There we can see a rich phase diagram as a function of $ma$ and $r$, particularly in the region away from $1 \gg ma$, where we are probing lattice-scale physics in the $4D$ theory. The purple region is the $\mathbb{Z}_{N\Gamma}$ symmetric phase; other colors denote less symmetric phases.  Note that at small values of $ma$, where we might hope the orbifold equivalence might hold and our 3D theory would correspond to a continuum 4D theory with fermions, the fermion contribution is unable to prevent the breaking of $\mathbb{Z}_\Gamma$ and the large N equivalence between the $3D$ and the $4D$ theory is lost. Even at $ma = 0$ and $r=0$, where the $4D$ theory would have  Dirac fermions with lattice doublers, which should help stabilize the center, the eigenvalues are in the totally collapsed phase, and $\mathbb{Z}_{\Gamma}$ is broken. A similar behavior was observed in one-loop calculations of effective potentials in matrix models of M-theory \cite{Kitsunezaki:1997iu}; the breaking of the symmetry there, like here, was  interpreted as a failure of the model to ``generate" spacetime dimensions. 

\section{Discussion}

The breaking of $\mathbb{Z}_\Gamma$ may seem at odds with similar calculations done in the past \cite{gross}, for instance the calculations  suggesting that adjoint fermions  protect orbifold equivalence and allow large $N$ dimensional reduction.  However, there is no contradiction: we would only expect the 3D effective potential to match the 4D effective potential for the trace of a Polyakov loop wrapping the compact direction if the two theories were large N  equivalent.  If the two theories are not equivalent, the 3D and 4D effective potentials do not have to agree even if they are effective potentials for the same order parameter.  As it happens, however, the effective potential that we have calculated in the $3D$ is for different order parameters than in the $4D$ calculations.

 In the 4D calculations in the literature,  $\mathop{\rm SU}(N)$ Yang-Mills theory\cite{seattle_vol_ind, gross, MyersOgilvie}, possibly with some matter content, was taken to be defined on $\mathbb{R}^3\times S^1$, with the size of the compactified dimension equal to $L$, and the calculations were done in the continuum limit.  To determine whether large $N$ volume-independence holds down to volumes small compared to the strong scale, the fate of the $\mathbb{Z}_N$ center symmetry was determined by looking at the effective potential for the Polyakov loop wrapping the compact spatial direction.  These calculations have shown that adjoint fermions protect the center symmetry at small volumes.   

Such 4D continuum calculations do not however answer the question of whether in a large N lattice theory with adjoint fermions (discretized along at least the compact direction), one can discard all but one lattice site along one direction and obtain a large N equivalent theory, which is the relevant question for determining whether orbifold equivalence holds and large N dimensional reduction works.  For that, one must examine the realization of the relevant discrete symmetries in the higher and lower dimensional theories, and it is precisely the latter question that our calculation of the effective potential in the 3D theory addresses.  Our results suggest that the relevant symmetries in the 3D theory break spontaneously, so that one does not obtain a large N equivalent theory in a full dimensional reduction even in the presence of light adjoint fermions.    

However, there remains a major puzzle.  There is a region in parameter space, which is colored purple in the phase diagram in Fig.~(\ref{fig:1}), where the full $Z_{N \Gamma}$ symmetry is unbroken, and orbifold equivalence appears to hold, at least at first glance: all of the conditions for large N orbifold equivalence appear to be met in this region of parameter space.  However, in this region the value of the fermion bare mass (equal at this order to the renormalized mass) is of the order of the cutoff $ma \sim 1$ and the fermions should effectively decouple from the low energy dynamics of the 4D theory.  The 4D theory should then be pure Yang-Mills in the continuum limit.  That is, again, contrary to any expectation.

To get further insight into our results, we can compare a quantity that can be calculated in both the 3D and 4D theories, which is the effective potential for a Wilson loop wrapping the compact direction in the $4D$ theory. For small enough $L\ll 1/\Lambda$, when it is legitimate to compute the effective potential in the semiclassical expansion, the effective potential can be calculated in both the 3D and 4D theories.  

In the pure Yang-Mills 4D theory, the effective potential for the Polyakov loop can be calculated by summing over the small fluctuations of the gauge field around a background gauge field with a non-trivial Polyakov loop:
\beq
A_\mu = \bar A_3 \delta^3_{\mu } + \mathcal{A}_\mu, \qquad
\bar A_3 = {\rm diag}( \theta_1/L ,\cdots, \theta_N/L ) , \qquad     
\Omega_3 = P\ e^{i\int dx^3\, \bar A_3 } =\begin{pmatrix}
                    e^{i\theta_1}    &                               &                 &     \\
                                          & e^{i\theta_2}            &                  &\\
                                          &                               & \ddots\\      &
                                          &                               &                 & e^{i\theta_N}  
                    \end{pmatrix}
\eeq 
The result, after gauge-fixing similar to Eq.~\eqref{eq:Veff} in $3D$, is \cite{seattle_vol_ind,gross,MyersOgilvie} 
\bea\label{eq:V4D}
V_{4D} &\sim& \frac{1}{L}\sum_{i,j=1}^{N} \sum_{n=1}^{\infty}\int d^3k\ \log\left(k^2+\left(\frac{2\pi n+\theta_i-\theta_j}{L} \right)^2\right)   \sim  - \frac{1}{L^4} \sum_{i,j=1}^{N} \sum_{n=1}^{\infty} \left(n+\frac{|\theta_i-\theta_j|}{2\pi}\right)^3 \nn\\
&\sim& - \frac{1}{L^4} \sum_{i,j=1}^N \zeta(-3, \frac{|\theta_i-\theta_j|}{2\pi})  
\sim   - \frac{1}{L^4} \sum_{i,j=1}^N\sum_{n=1}^{\infty} \frac{\cos(n(\theta_i-\theta_j))}{n^4} \nn  \\
&\sim&    - \frac{1}{L^4} \sum_{n=1}^{\infty} \frac{1}{n^4}|{\tr }\ \Omega_{3}^{n}|^2 ,
\eea 
where we used the $\zeta$-function regularization to define the sum over the Kaluza-Klein modes, and used Hurwitz's formula in the next to last equality.  The result can be written in two useful ways \cite{seattle_vol_ind}, one making the sum over the winding numbers of the Polyakov loop explicit and the other making the sum over the eigenvalues explicit:
\bea
\label{eq:Veff_4D}
V_{4D} &=&    - \frac{2}{\pi^2 L^4} \sum_{n=1}^{\infty} \frac{1}{n^4}|{\tr }\ \Omega_{3}^{n}|^2 \nn \\
V_{4D} &=&    - \frac{1}{24 \pi^2L^4}\left(\frac{8\pi^4 N^2}{15} - \sum_{i,j=1}^{N}[\theta_i - \theta_j]^2([\theta_i - \theta_j]-2\pi^2)^2  \right) \;.
\eea 
where $[a] \equiv a \mod 2\pi$.  It is clear from these expression that the potential is minimized when all eigenvalues collapse to one common value.  Each flavor of adjoint Majorana fermions makes an equal and opposite contribution to that of the gauge bosons to the effective potential, so that adding one flavor of Dirac fermions, or equivalently two flavors of Majorana fermions, will simply have the effect of flipping the sign of the effective potential above, making it repulsive.  The resulting potential favors $\mathbb{Z}_{N}$ symmetric distributions of eigenvalues of the Polyakov loop.

We can also calculate the effective potential for the Polyakov loop from the 3D theory.  In fact, the calculation of the effective potential we have done above is general enough that it already gives the answer in the appropriate sector.  If we consider a $\phi_0$ that takes the block-diagonal form
\beq\label{eq:phi0_blocked}
\phi_0 = \begin{pmatrix}
e^{i\varphi_i} \mathbf{1}_\Gamma &  & \\
&    \ddots &  \\
&       &  e^{i\varphi_N} \mathbf{1}_\Gamma 
\end{pmatrix},
\eeq 
where $\mathbf{1}_{\Gamma}$ is the $\Gamma \times \Gamma$ identity matrix,  $\Gamma \varphi_i = \theta_i$, then $\phi^\Gamma=\Omega_3$ and, in the region of parameter space where the large N equivalence holds, the effective potentials for the eigenvalues of $\Omega_3$ calculated in the 4D and 3D theories should be the same. Let us look at the effective potential for $\tr \; \Omega_3$ as calculated from the 3D theory by making making the appropriate substitutions in Eq.~(\ref{eq:Veff}). In the continuum limit, where $\Gamma \rightarrow \infty, a \rightarrow 0$, with $\Gamma a =L $ fixed, Eq.~(\ref{eq:Veff}) reduces to 
\beq
V_{\mathrm{eff}}(\Omega_3) = \frac{1}{6 \pi L^3} \sum_{i<j}^N\left( \sqrt{2} |\theta_i-\theta_j|^3 -(8 (m L)^3 +12 m L (a m r - 1)|\theta_i-\theta_j |^2 )\right) .
\eeq 
Note that the term with the Wilson term coefficient disappears in the continuum limit of $a\rightarrow 0$, as it must.  However, this effective potential does not match Eq.~(\ref{eq:Veff_4D}) for any value of $m$.  In particular, the two results clearly do not match for $m \approx 0.7/a, r \approx 1 $, which is in the region in parameter space where $\mathbb{Z}_{\Gamma}$ symmetry is preserved, and large N orbifold equivalence would naively be expected to hold.   Probably the most important difference between the two effective potentials is the lack of anything like a sum over Kaluza-Klein modes in the $3D$ theory, while such a sum appears naturally in the $4D$ calculation of the effective potential. 

 In our calculation of the effective potential of the $3D$ theory, we took  the scalar field $\phi$ to be close to a block diagonal form.  But it was the one-off-diagonal terms in Eq.~(\ref{eq:projected}) that generated the derivative terms in the ``extra" dimension. This means that we had essentially no chance to see derivatives of the KK modes arise in our computation, and thus each of the terms in the sum over the KK modes is independent of the KK mode number $n$.  The sum over $n$ reduces in our case to a factor of $\Gamma$, with the $|\theta_i-\theta_j|^3$ dependence agreeing between the 4D and the 3D calculation.  Of course, if large N orbifold equivalence were to hold, the behavior of gauge-invariant observables in the common sector of the two theories {\it should} agree regardless.  The fact that the effective potential for center symmetry breaking calculated in the 4D theory and in the 3D theory do not agree even in the region of parameter space where $\mathbb{Z}_{\Gamma}$ symmetry is unbroken suggests that the equivalence is breaking down for some other reason that is not currently understood.
 
One way to reconcile the continuum 4D calculations with our construction is the following\footnote{We thank M. Unsal for conversations on this point.}. The 4D calculations suggest that $L$ can be reduced arbitrarily as long as $L\gg a$, that is, within the region of validity of a continuum approximation. A theory like the 3D theory described here can be seen as the result of reducing $L$ all the way to $L=a$ (even though what we consider in this paper is dimensional expansion from 3 to 4 dimensions and $L =\Gamma a\gg a$ at all times). From that point of view, the results stemming from our 3D theory are lattice artifacts. That would suggest that a YM theory with adjoint fermions may have large N volume independence which works for arbitrarily small $L$ {\it in physical units}, but the resulting volume-reduced theory must still have a large number of lattice points in the compactified direction, which look like a large number of independent fields form the 3D point of view. Viewing such a ``reduced" theory as being 3D does not seem to provide any advantage over just working with the original 4D theory.

The challenge of unraveling this situation is left for future work.  A working orbifold equivalence allowing large N equivalence between interesting theories such as Yang-Mills theory (with matter or without) in different numbers of spacetime dimensions would be extremely useful.  Unfortunately, as we have seen in this paper, it is not yet clear how to make such an equivalence work.

\acknowledgments 
The work of P.~F.~B., M.~I.~B, and A.~C. was supported by the US Department of Energy through grant DE-FG02-93ER-40762, and the work of R.~P.~S. was supported by the US Department of Energy through grant DE-FG02-05ER41368.   We thank Tom Cohen, Barak Bringoltz, Herbert Neuberger, Michael Ogilvie, Brian Tiburzi, Mithat Unsal, and Larry Yaffe for enlightening conversations, and particularly thank Mithat Unsal and Larry Yaffe for a critical reading of the manuscript.  M.~I.~B and A.~C. thank the Institute for Nuclear Theory and the organizers and participants of the workshop ``New Frontiers in Large N Gauge Theories" where this work was discussed.   

\section{Postscript}
After the completion of the present work and partially triggered by it, Bringoltz\cite{barack} analyzed similar questions from a different perspective. His analysis sheds some light on the issues we discuss above, and we would like to comment on it here.

Ref.~\cite{barack} pointed out that the 3D reduced model in Eq.~(\ref{eq:3d}) is non-renormalizable in perturbation theory, and that the linear divergences that arise in calculating the one-loop effective potential for $\mathbb{Z}_{N\Gamma}$ breaking
\footnote{Refs.~\cite{barack, neuberger_private} note that if the ``extra" dimension is shrunk to two or  more points (instead of one), the linear divergence we discussed here disappears. This indicates that a continuum 3D model with {\it two} unitary adjoint scalars should not have the linear divergence we are discussing. This is essentially an example of the Little Higgs construction \cite{little_higgs} popular in electroweak symmetry breaking model building. }  
must  be absorbed by counterterms coming from certain relevant operators that should be added to the action in Eq.~(\ref{eq:3d}) and not, as we erroneously assumed before, by terms already contained in the action.
For instance, in the bosonic sector the effective potential given in Eq.~(\ref{eq:integral}) evaluated with a ultraviolet cutoff    $\Lambda$ is proportional to
\beq
\frac{4\pi}{(2\pi)^3} \int^{\Lambda}{dk\, k^2 \log(k^2-M^2)} =-\frac{\Lambda M^2}{2\pi^2} - \frac{i M^3}{6\pi}+\mathcal{O}(\Lambda^{-1}),
\eeq
where $M^2 = \frac{2}{a^2}\sin^2\left(\frac{\varphi_i-\varphi_j}{2}\right)$, and we have discarded a field-independent divergent constant.  The linear divergence is absorbed by
\beq
 d_1 \int d^3x |{\rm tr}\phi|^2 
\eeq 
and not, as previously assumed, by a wave function renormalization. When fermions are included two counterterms are required at one-loop
\beq
\label{eq:double-trace}
 d_1 \int d^3x |{\rm tr}\phi|^2 +  d_2 \int d^3x |{\rm tr}\phi^2|^2 \;.
\eeq 
It seems then that an interesting question is whether the theory in Eq.~(\ref{eq:3d}) supplemented by these two terms can be equivalent to a four-dimensional gauge theory. For this to be true two conditions need to be met. First, $\mathbb{Z}_{N \Gamma}$ must not break. Secondly, the theory should be orbifold projected into a theory equivalent to the four-dimensional gauge theory of interest. 

\begin{figure}[h]
\begin{center}$
\begin{array}{ccc}
\includegraphics[scale=0.9]{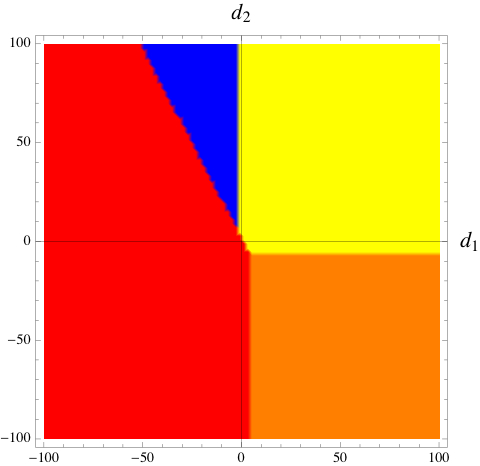} &
\includegraphics[scale=0.9]{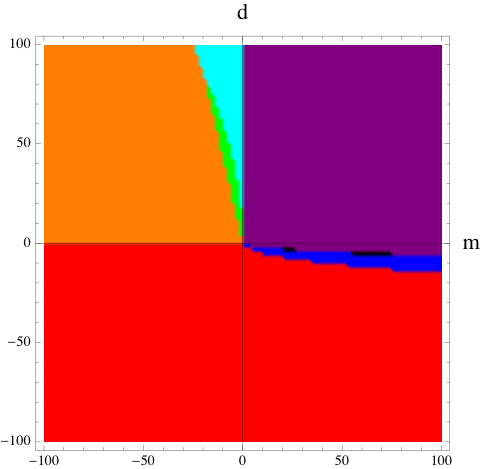} &
\includegraphics[scale=0.35]{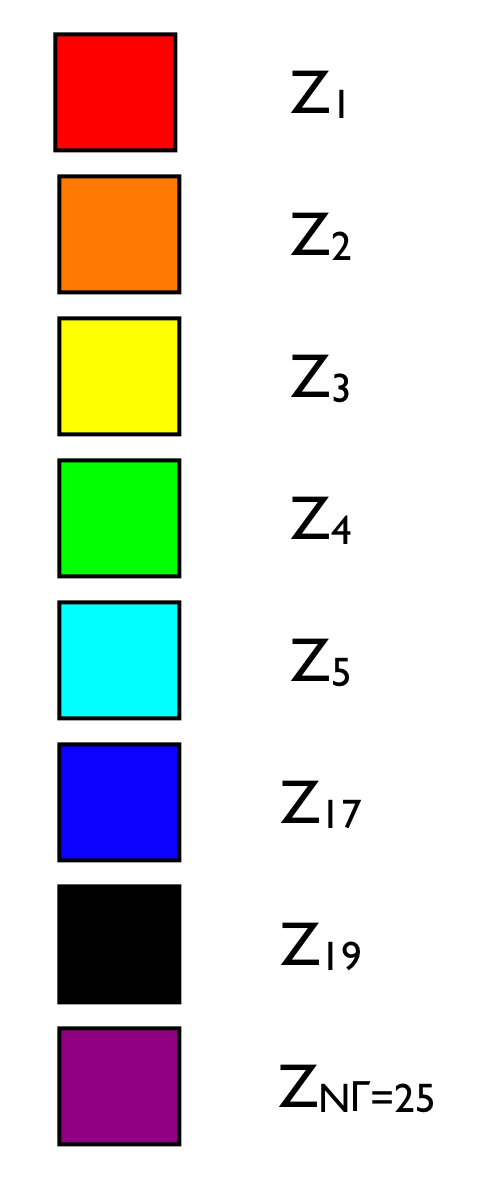} 
\end{array}$
\end{center}
\caption{The figure on the left shows the $\mathbb{Z}_{N\Gamma}$ phase diagram following from numerical minimization of the one-loop effective potential with only the first two double-trace terms included with coefficients $d_1, d_2$.  The figure on the right shows the phase diagram with the full tower of double trace terms, with all double trace terms having the same coefficient $d$, with varying bare fermion mass $m$ and Wilson term coefficient $r=1$.  In both figures $N\Gamma=25$ and $N_f=1$.     }
\label{fig:DTs}
\end{figure}

The coefficients $d_1, d_2$ contain regulator-dependent divergent pieces, which serve to cancel the one-loop UV divergences, and also contain arbitrary finite pieces that contribute to the effective potential at tree level.  The values of the finite pieces can affect the realization of $\mathbb{Z}_{N \Gamma}$ symmetry.   Thus, to check the first condition above, we added the two double trace terms in Eq.~(\ref{eq:double-trace}) and varied their coefficients while checking the center symmetry realization using the methods used above in Sec.~\ref{sec:PhaseDiagram}.   We were \emph{unable}\footnote{The exception is if $N\Gamma$ is prime and one allows the double-trace term coefficients to scale with $N \Gamma$, which is not what one wants to do for large $N$ equivalence.} to find double-trace term coefficients that would protect center symmetry in the theory with $N_f \neq 0$.  As is illustrated in Fig.~\ref{fig:DTs} for $N_f=1$, the center symmetry  breaks to $\mathbb{Z}_3$, even when the UV-finite parts of the double trace coefficients are large\footnote{The situation for $N_f=0$ is somewhat different at one loop, since a direct numerical check shows that the two double trace terms in Eq.~(\ref{eq:double-trace}) can indeed prevent center symmetry breaking.  However, we do not expect this to hold beyond the one-loop approximation, and the full tower of double trace terms is likely to be necessary to protect center symmetry beyond one loop even for $N_f=0$.}.  

This result is expected by general considerations. The stabilization of center symmetry using double trace terms was discussed extensively in Ref.~\cite{doubletrace}.  As argued there, it is expected that if the coefficients of the terms in Eq.~(\ref{eq:double-trace}) are chosen appropriately, one can prevent center symmetry breaking with $\tr \phi$ and $\tr \phi^{2}$ as order parameters.  However, no matter what their coefficients are, these two double trace terms would not generally be expected to prevent the breaking of center symmetry with higher powers of $\phi$ as order parameters, for instance breaking with $\tr \phi^3$ as an order parameter.  (This is breaking pattern referred to as $\mathbb{Z}_3$ in Sec.~\ref{sec:PhaseDiagram}). As argued by Unsal and Yaffe\cite{doubletrace}, to fully protect center symmetry using double trace deformations, one would need to include a whole tower of double trace terms, with all powers of $\phi$ up to $N\Gamma/2$:
\beq
\label{eq:AllDoubleTraces}
\sum_{k=1}^{N\Gamma/2 }d_k \int d^3x |{\rm  tr}\phi |^{2k} \;.
\eeq 
This suggests the addition of the full tower of double-trace operators to the action, whether or not their presence is required by renormalization at one-loop.   Once the full tower of double trace terms in Eq.~(\ref{eq:AllDoubleTraces}) is included, center symmetry can easily be protected by choosing appropriate coefficients for the double-trace terms, as is illustrated in Fig.~\ref{fig:DTs}~\footnote{Interestingly, one can also preserve center symmetry by putting the full tower of double trace terms except for all terms that contain  $\phi^{k\Gamma}$, where $k$ are integers.  This suggests that one can protect the center symmetry with a double trace deformation that does not survive the orbifold projection. The implications of this curious result are left to future work.} 

We now turn to addressing the second condition: whether the theory with the added double-trace operators orbifolds into a theory equivalent to a four-dimensional gauge theory.
It can be argued that the theory with the added double-trace terms and $d_k \sim 1/L^3=1/(\Gamma a)^3$ is equivalent to a theory without them on two accounts in the large volume limit. First, in the large  $L$ limit the deformation vanishes. Second, as argued Ref.~\cite{doubletrace},  their contribution to the dynamics at large volume in the large $N$ limit are suppressed by $1/N^2$. In our case, the  double-trace terms orbifold into
\beq
\label{eq:orbifolded-double-trace}
d_{n\Gamma} \int{d^{3}x\, |\tr \phi^{n \Gamma}|^2} \rightarrow \Gamma d_{n\Gamma}  \int{d^{3} x\, \tr |\Omega^n|^2}
\eeq 
where $\tr \Omega = \tr \phi^{\Gamma}$ is a traced Wilson loop wrapping the compact direction in the 4D theory. Recall that terms like $|\tr\phi^{p}|^2, p \notin \{m \Gamma, m \in \mathbb{N} \}$ do not survive the volume-expanding orbifold projection.  Thus it is clear that the two double trace terms whose coefficients contain regulator-dependent UV divergent pieces at one loop do not appear in the orbifolded theory.  Also, it is clear that at one loop, the coefficients $d_k$ with $k>2$, some of which survive the orbifold, are independent of the regulator of the 3D theory.  Since we want $d_{k}$ to make contributions to the one-loop effective potential of comparable size to the other terms in the action (that is, finite in the large $\Gamma$ and large $N$ limits), we must generically take $d_{k}= f_k/a^3$, where $f_k$ are numbers independent of $\Gamma$ and $N$. 

The resulting four-dimensional double-trace  terms clearly do not vanish in the large $L = \Gamma a$ limit, and the first argument given above for the irrelevance of the double-trace terms in the large-volume 4D theory no longer applies.  However, the second argument, involving the $1/N^2$ suppression, will still work provided that the large $N$ limit is taken \emph{before} the continuum limit and before the large volume limit: the limits do not commute.   With this ordering of limits, once the values of coefficients $d_k$ are chosen appropriately, the 3D theory will be orbifold-equivalent at large $N$ to 4D YM with adjoint fermions with double trace deformations, and at large volume $L$ and large $N$, this latter theory will be equivalent to the desired 4D theory, which is simply 4D YM with adjoint fermions.

 The upshot of the above discussion is that to prevent center symmetry breaking and have a 3D theory equivalent to a 4D gauge theory in the large $N$ limit it is necessary to add the full tower of double-trace operators to the three-dimensional action and rely on the $1/N^2$ suppression of the double-trace operators in the large $N$ limit.

Ref.~\cite{barack} approached the problem not from a continuum theory as we did in this paper but worked with a lattice version\footnote{See also Ref.~\cite{barack2} for a related numerical study of a reduction of the 4D theory to one plaquette. } of Yang-Mills with adjoint Wilson fermions with a variable number of sites.  When the number of sites in all directions is large, taking the continuum limit one obtains continuum 4D Yang-Mills theory with adjoint fermions, plus higher dimensional operators.  On the other hand, when the number of sites in the three non-compact directions is large, but there is only one link in the compact direction, the continuum limit gives the theory discussed in the present paper, with the link in the compact direction denoted by $\phi$, supplemented by the double-trace operators in Eq.~(\ref{eq:double-trace}).   In the construction of Ref.~\cite{barack} there are not explicit double-trace operators in the action. However, if one desires to discuss the continuum limit of the lattice theory, one can view our continuum theory as the low energy effective theory for it. From this perspective, the numerical values of $d_1,d_2$ in the effective theory are determined from the lattice theory, and {\it are dependent on the particular lattice regularization used}.  Working in perturbation theory it was shown that two double-trace operators are indeed generated at one-loop level. As argued before, they are not enough to suppress $\mathbb{Z}_{N\Gamma}$ breaking.  

At first glance, it might appear that there is some tension between these results and that of Ref.~\cite{barack}.  In that work, $\mathbb{Z}_{N\Gamma}$ breaking was tested in the lattice-regularized theory with adjoint fermions (which assigns some particular values to the double trace coefficients in Eq.~(\ref{eq:double-trace})), and it was argued that $\mathbb{Z}_{N\Gamma}$ did not break at one loop.  This conclusion was based on comparing the potential energy of the  $\mathbb{Z}_1,\mathbb{Z}_2$, and $\mathbb{Z}_{N\Gamma}$ symmetry realizations, and the $\mathbb{Z}_{N\Gamma}$ symmetric realization was found to be preferred to the other two tested.  If, as suggested by our results above, the center symmetry in the lattice-regulated theory breaks with $\tr \phi^3$ with an order parameter, this would not have shown up in the analysis of Ref.~\cite{barack}.  If, on the other hand, $\mathbb{Z}_{N\Gamma}$ breaking does not occur after all in the lattice-regulated theory, consistency with our results above would suggest that the lattice-regulated theory with fermions actually generates the whole double-trace tower \emph{at one loop} in the continuum limit, rather than just the first two terms considered in Ref.~\cite{barack}.

We now finally return to the question raised in the title of this paper:  do fermions save large N dimensional reduction?  Working with the 3D continuum theory, which is the choice we investigated here, the answer is no.  The preservation of center symmetry depends \emph{entirely} on the coefficients of the double-trace terms in the action, which generically have UV-divergent pieces that depend on the choice of regulator, and finite parts that can be chosen freely by hand from the EFT point of view.  We see that while it appears that (happily) large $N$ dimensional reduction \emph{can} be saved, fermions do not participate in the rescue:  The fermions themselves are not enough to protect center symmetry of the 3D theory, and just turning on adjoint fermions is not enough to make the 3D and 4D theories orbifold equivalent.   

The answer to the same question is more subtle in the lattice version of the 3D theory which was investigated in Ref.~\cite{barack}.   There, it was argued that the lattice regulator of the theory with Wilson fermions effectively assigns certain UV-divergent values to the coefficients of the first two double-trace terms that turn out to be sufficient to protect center symmetry in the 3D lattice theory with one site in the ``extra'' direction.  If this is indeed the case, the center symmetry of the 3D lattice theory is saved by UV effects associated with the fermion discretization, rather than the infrared physics of the fermions, as might have been expected given the results found in the analogous 4D continuum theory in Ref.~\cite{seattle_vol_ind} with adjoint fermions in small volumes.


\end{document}